\documentclass[a4paper,10pt]{amsart}
 \usepackage{graphicx,amsfonts,amsmath,amssymb,latexsym,cancel}

  \theoremstyle{plain}
 \newtheorem {hypo}{\bf\hspace{-\parindent}Hypothesis}

 \newtheorem {thmx}[hypo]{Theorem}
 \newtheorem {prop}[hypo]{Proposition}
 
 \newtheorem {conj}[hypo]{Conjecture}
 
 \newtheorem {lem}[hypo]{Lemma}

 \theoremstyle{definition}
 \newtheorem {eg}[hypo]{Example}

 \theoremstyle{remark}
 \newtheorem {rmk}[hypo]{Remark}

 \newcommand\Zb{\mathbb{Z}}
 
 \newcommand\Cb{\mathbb{C}}
 \newcommand\Pb{\mathbb{P}}

 \newcommand\ben{\begin{equation*}}
 \newcommand\ebn{\end{equation*}}
 \newcommand\be{\begin{equation}}
 \newcommand\eb{\end{equation}}
 \newcommand\ds{\displaystyle}
 
 \newcommand{\pf}{\begin{bpf}}

 \newcommand{\pfms}{\begin{bpfms}}
 \newcommand{\epf}{\end{bpf}\hfill$\square$\vspace{0.1cm}}
 \newcommand{\epfms}{\end{bpfms}\hfill$\square$\\ }

 \numberwithin{equation}{section}

 \title{Dyson's constant for the hypergeometric kernel}
 \author{Oleg Lisovyy}
 \address{Laboratoire de Math\'ematiques et Physique Th\'eorique CNRS/UMR 6083,
 Universit\'e de Tours, Parc de Grandmont, 37200 Tours, France}
 \email{lisovyi@lmpt.univ-tours.fr}
 \address{Bogolyubov Institute for Theoretical Physics, 03680 Kyiv, Ukraine}
 \date{}

\begin{document}
 \begin{abstract}
 We study a Fredholm determinant of the hypergeometric kernel arising in
 the representation theory of the infinite-dimensional unitary group. It is shown
 that this determinant coincides with the Palmer-Beatty-Tracy
 tau function of a Dirac operator on the hyperbolic disk.
 Solution of the connection problem for Painlev\'e~VI equation allows to determine
 its asymptotic behavior up to a constant factor, for which a conjectural expression
 is given in terms of Barnes functions. We also present analogous asymptotic results
 for the Whittaker and Macdonald kernel.
 \end{abstract}

 \maketitle

  \section{Introduction}
  Connections between Painlev\'e equations and Fredholm determinants have long been
  a subject of great interest, mainly because of their applications in
  random matrix theory and integrable systems, see~e.g. \cite{jmms,twreview,wmtb}. One
  of the most famous examples is concerned with the Fredholm determinant
  $F(t)=\mathrm{det}(1-K_{\text{sine}})$, where $K_{\text{sine}}$ is the integral operator with the sine
  kernel $\frac{\sin(x-y)}{\pi(x-y)}$ on the interval $[0,t]$.  It is well-known that $F(t)$
  is equal to the gap probability for the Gaussian Unitary Ensemble (GUE)
  in the bulk scaling limit. As shown in \cite{jmms}, the function $\ds \sigma(t)=t\frac{d}{dt}\ln F(t)$
  satisfies the $\sigma$-form of a Painlev\'e~V equation,
  \be\label{pvsine}
  \left(t\sigma''\right)^2+4\left(t\sigma'-\sigma\right)\bigl(t\sigma'-\sigma+(\sigma')^2\bigr)=0.
  \eb

  Equation (\ref{pvsine}) and the obvious leading behavior  $F(t\rightarrow0)=1-t+O\left(t^2\right)$
  provide an efficient method of numerical computation of $F(t)$ for all $t$. Further,
  as $t\rightarrow\infty$, one has
  \ben
  F(2t)=f_0\,t^{-\frac14}e^{-t^2/2}\Bigl(1+\sum\limits_{k=1}^{N}f_k t^{-k}+O\bigl(t^{-N-1}\bigr)\Bigr).
  \ebn
  The coefficients $f_1,f_2,\ldots$ in this expansion can in principle be determined from (\ref{pvsine}).
  It was conjectured by Dyson \cite{dyson} that the value of the remaining unknown constant is
  $\ds f_0=2^{\frac{1}{12}}e^{3\zeta'(-1)}$, where $\zeta(z)$ is the Riemann $\zeta$-function.

  Dyson's conjecture was rigorously proved only recently in \cite{dikz,ehrhardt,krasovsky}.
  Similar results were also obtained in \cite{baik,dik} for the Airy-kernel determinant
  describing the largest eigenvalue distribution for GUE in the edge scaling limit \cite{twairy}.

  The present paper is devoted to the asymptotic analysis of the Fredholm determinant
  of the hypergeometric kernel on $L^2(0,t)$ with $t\in(0,1)$. This determinant,
  to be denoted by $D(t)$,  arises in the representation theory of the infinite-dimensional
  unitary group \cite{BO5} and provides a 4-parameter class of solutions
  to Painlev\'e~VI (PVI) equation \cite{borodin}. Rather surprisingly,
  it turns out to coincide with the Palmer-Beatty-Tracy (PBT) $\tau$-function of
  a Dirac operator on the hyperbolic disk \cite{lisovyy, beatty} under suitable identification
  of parameters. Relation to PVI allows to give a complete description of the behavior
  of $D(t)$ as $t\rightarrow1$ up to a constant factor analogous to Dyson's constant $f_0$
  in the sine-kernel asymptotics. Relation to the PBT $\tau$-function, on the other hand,
  suggests a conjectural expression for this constant in terms of Barnes functions.

  The paper is planned as follows. In Section~2, we recall basic facts on Painlev\'e~VI
  and the associated linear system. The $_2F_1$ kernel determinant $D(t)$ and the PBT $\tau$-function
  are introduced in Sections~\ref{f21section} and~\ref{pbtsection}.
  Section~\ref{twsection} gives a simple proof of a result of \cite{borodin}, relating $D(t)$ to Painlev\'e~VI.
  In Section~\ref{jafsection},
  we discuss Jimbo's asymptotic formula for PVI and determine the monodromy corresponding
  to the $_2F_1$ kernel solution. Section~\ref{mainsection} contains the main results of the paper:
  the asymptotics of $D(t)$ as $t\rightarrow1$, obtained from the solution of PVI connection problem
  (Proposition~\ref{UVas}) and a conjecture for the unknown constant (Conjecture~\ref{dysonconj}).
  Numerical and analytic tests of the conjecture are discussed in Sections~\ref{nchecksection}
  and~\ref{achecksection}. Similar asymptotic results for the Whittaker and Macdonald kernel
  are presented in Section~\ref{wmsection}. Appendix~A contains a brief summary
  of formulas for the Barnes function.

 \section{Painlev\'e VI and JMU $\tau$-function}\label{pvisection}
 Consider the linear system
 \be\label{fuchsian}
 \frac{d\Phi}{d\lambda}=\left(\frac{A_0}{\lambda}+\frac{A_1}{\lambda-1}+
 \frac{A_t}{\lambda-t}\right)\Phi,
 \eb
 where $A_{\nu}\in\mathfrak{sl}_2(\Cb)$ ($\nu=0,1,t$) are independent of $\lambda$
 with eigenvalues $\pm \theta_{\nu}/2$  and
 \ben
 A_0+A_1+A_t=\left(\begin{array}{cc}-\theta_{\infty}/2 & 0 \\ 0 &
 \theta_{\infty}/2\end{array}\right), \qquad\theta_{\infty}\neq0.
 \ebn
 The fundamental matrix solution $\Phi(\lambda)$ is a multivalued function on
 $\Pb^1\backslash\{0,1,t,\infty\}$. Fix the basis of loops as shown in Fig.~1 and denote
 by $M_0,M_t,M_1,M_{\infty}\in SL(2,\Cb)$ the corresponding monodromy matrices.
 Clearly, one has $M_{\infty}M_1M_tM_0=\mathbf{1}$.
 \begin{figure}[!h]
 \begin{center}
 \resizebox{5cm}{!}{
 \includegraphics{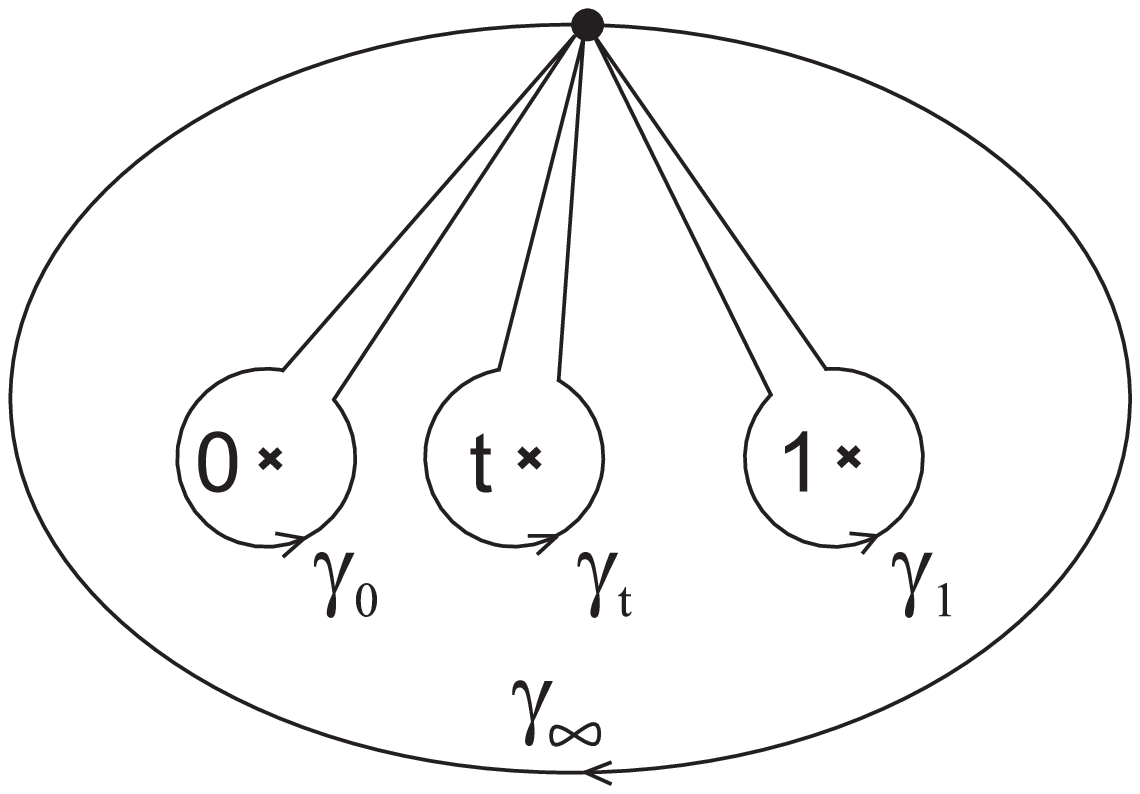}} \\
 Fig. 1: Generators of $\pi_1\left(\Pb^1\backslash\{0,1,t,\infty\}\right)$.
 \end{center}
 \end{figure}

 Since the monodromy is defined up to overall conjugation, it is convenient to introduce,
 following \cite{jimbo}, a 7-tuple of invariant quantities
 \begin{align}
 \label{pnu}
 p_{\nu}=&\;\mathrm{Tr}\,M_{\nu}=2\cos\pi\theta_{\nu},\qquad\qquad \nu=0,1,t,\infty,
 \\
 \label{pmunu}
 p_{\mu\nu}=&\;\mathrm{Tr}\left(M_{\mu}M_{\nu}\right)=2\cos\pi\sigma_{\mu\nu},\qquad \mu,\nu=0,1,t.
 \end{align}
 These data uniquely fix the conjugacy class of the triple $(M_0,M_1,M_t)$ unless the
 monodromy is reducible. The traces (\ref{pnu})--(\ref{pmunu}) satisfy Jimbo-Fricke relation
 \ben
 p_{0t}p_{1t}p_{01}+p_{0t}^2+p_{1t}^2+p_{01}^2-(p_0p_t+p_1p_{\infty})p_{0t}-
 (p_1p_t+p_0p_{\infty})p_{1t}-(p_0p_1+p_tp_{\infty})p_{01}=4.
 \ebn
 As a consequence, for fixed $\{p_{\nu}\}$, $p_{0t}$, $p_{1t}$ there are at most two possible
 values for $p_{01}$.

 It is well-known that the monodromy preserving deformations of the system (\ref{fuchsian})
 are described by the so-called Schlesinger equations
 \be\label{schs}
 \frac{dA_0}{dt}=\frac{[A_t,A_0]}{t},\qquad
 \frac{dA_1}{dt}=\frac{[A_t,A_1]}{t-1},
 \eb
 which are equivalent to the sixth Painlev\'e equation:
 \begin{align}
 \label{pvi}\frac{d^2q}{dt^2}=\;\frac{1}{2}\left(\frac{1}{q}+\frac{1}{q-1}
 +\frac{1}{q-t}\right)\left(\frac{dq}{dt}\right)^2 -
 \left(\frac{1}{t}+\frac{1}{t-1}+\frac{1}{q-t}\right)\frac{dq}{dt}\,+\\
 \nonumber +\,
 \frac{q(q-1)(q-t)}{2t^2(t-1)^2}\left((\theta_{\infty}-1)^2-\frac{\theta_0^2 t}{q^2}+
 \frac{\theta_1^2(t-1)}{(q-1)^2}+\frac{(1-\theta_t^2)t(t-1)}{(q-t)^2}\right).
 \end{align}
 Relation between $A_{0,1,t}(t)$ and  $q(t)$ is given by
 $\text{\small $\ds
 \left(\frac{A_0}{\lambda}+\frac{A_1}{\lambda-1}+
 \frac{A_t}{\lambda-t}\right)_{12}=\frac{k(t)(\lambda-q(t))}{\lambda(\lambda-1)(\lambda-t)}.
 $}$

  Jimbo-Miwa-Ueno (JMU) $\tau$-function \cite{jmu} of Painlev\'e~VI is defined as follows:
 \be\label{jmutau}
 \frac{d}{dt}\ln\tau_{\scriptscriptstyle JMU}(t;\boldsymbol{\theta})=
 \frac{\mathrm{tr}\left(A_0A_t\right)}{t}+\frac{\mathrm{tr}\left(A_1A_t\right)}{t-1}\,,
 \eb
 where $\boldsymbol{\theta}=(\theta_0,\theta_1,\theta_t,\theta_{\infty})$.
 Introducing a logarithmic derivative
 \be\label{zetatau}
 \sigma(t)=t(t-1)\frac{d}{dt}\ln\tau_{\scriptscriptstyle JMU}(t;\boldsymbol{\theta})+
 \frac{t\left(\theta_t^2-\theta_{\infty}^2\right)}{4}-
 \frac{\theta_t^2+\theta_0^2-\theta_1^2-\theta_{\infty}^2}{8},
 \eb
 it can be deduced from the Schlesinger system (\ref{schs}) that $\sigma(t)$ satisfies
 the following 2nd order ODE ($\sigma$-form of Painlev\'e~VI):
  \begin{align}
 \label{zetapvi}
 \sigma'\Bigl(t(t-1)\sigma''\Bigr)^2+\left[2\sigma'(t\sigma'-\sigma)-\left(\sigma'\right)^2-
 \frac{(\theta_t^2-\theta_{\infty}^2)(\theta_0^2-\theta_1^2)}{16}\right]^2= \\
 \nonumber =\,
 \left(\sigma'+\frac{(\theta_t+\theta_{\infty})^2}{4}\right)
 \left(\sigma'+\frac{(\theta_t-\theta_{\infty})^2}{4}\right)
 \left(\sigma'+\frac{(\theta_0+\theta_{1})^2}{4}\right)
 \left(\sigma'+\frac{(\theta_0-\theta_{1})^2}{4}\right).
 \end{align}
 In terms of $q(t)$, the definition of $\sigma(t)$ reads
 \begin{align}\label{zeta1}
 \sigma(t)=&\,\frac{t^2 (t-1)^2}{4q(q-1)(q-t)}\left(q'-\frac{q(q-1)}{t(t-1)}\right)^2-
 \frac{\theta_0^2 t}{4q}+\frac{\theta_1^2(t-1)}{4(q-1)}-\frac{\theta_t^2 t(t-1)}{4(q-t)}\\
 \nonumber &\, -\frac{\theta_{\infty}^2(q-1)}{4}-\frac{\theta_t^2 t}{4}
 +\frac{\theta_t^2+\theta_0^2-\theta_1^2-\theta_{\infty}^2}{8}.
 \end{align}

 \section{Hypergeometric kernel determinant}\label{f21section}
 It was shown in \cite{BO5} that the spectral measure associated to the decomposition
 of a remarkable 4-parameter family of characters of the infinite-dimensional
 unitary group $U(\infty)$ gives rise to a determinantal point process with
 correlation kernel
  \ben
  K(x,y)=\lambda\;\frac{A(x)B(y)-B(x)A(y)}{y-x},\qquad x,y\in(0,1),
  \ebn
  where
  \begin{align}
  \label{f21lambda}
  \lambda\,=&\;\frac{\sin\pi z\sin\pi z'}{\pi^2}\,
  \Gamma\left[\begin{array}{c}1+z+w,1+z+w',1+z'+w,1+z'+w' \\ 1+z+z'+w+w',2+z+z'+w+w'\end{array}\right],\\
  \label{f21A}
  A(x)=&\;x^{\frac{z+z'+w+w'}{2}}\left(1-x\right)^{-\frac{z+z'+2w'}{2}}
   {}_2F_1\left[\left.
  \begin{array}{c}z+w',z'+w' \\ z+z'+w+w' \end{array}\right|\frac{x}{x-1}\right],\\
  \label{f21B}
  B(x)=&\;x^{\frac{z+z'+w+w'+2}{2}}\left(1-x\right)^{-\frac{z+z'+2w'+2}{2}}
  {}_2F_1\left[\left. \begin{array}{c}z+w'+1,z'+w'+1 \\ z+z'+w+w'+2 \end{array}\right|\frac{x}{x-1}\right].
  \end{align}
  Note that our notation slightly differs from the standard one \cite{BO5,borodin};
  to shorten some formulas from Painlev\'e theory, the interval $\left(\frac12,\infty\right)$
  of \cite{BO5,borodin} is mapped to $(0,1)$ by $x\mapsto 1/{\left(\frac12+x\right)}$.

  The kernel $K(x,y)$ has a number of symmetries:
  \begin{enumerate}
  \item[(S1)]
  It is invariant under transformations $z\leftrightarrow z'$ and
  $w\leftrightarrow w'$; the latter symmetry follows from
  $  _2F_1\left[\text{\footnotesize$\begin{array}{c} a,b\\ c \end{array}$}\bigl|\,z\right]=
  (1-z)^{c-a-b} {}_2F_1\left[\text{\footnotesize$\begin{array}{c} c-a,c-b\\ c \end{array}$}\bigl|\,z\right]$.
  \item[(S2)]
  It is also straightforward to check that $K(x,y)$ is invariant under
  transformation
  \ben
  \text{$z\mapsto -z$, $z'\mapsto-z'$, $w\mapsto w'+z+z'$, $w'\mapsto w+z+z'$}.
  \ebn
  \item[(S3)] We can simultaneously shift
  $\text{$z\mapsto z\pm1$, $z'\mapsto z'\pm1$, $w\mapsto w\mp1$, $w'\mapsto w'\mp1$}$;
  toge\-ther with (S2), this allows to assume without loss of generality that
  $0\leq\mathrm{Re}\left(z+z'\right)\leq1$.
  \end{enumerate}

 We are interested in the Fredholm determinant
  \be\label{f21det}
  D(t)=\mathrm{det}\left(1-K\bigl|_{\left(0,t\right)}\right),\qquad t\in \left(0,1\right).
  \eb
 Assume that the parameters $z,z',w,w'\in\Cb$ satisfy the conditions:
  \begin{enumerate}
  \item[(C1)] $z'=\bar{z}\in\Cb\backslash\Zb$ or $k<z,z'<k+1$ for some $k\in\Zb$,
  \item[(C2)] $w'=\bar{w}\in\Cb\backslash\Zb$ or $l<w,w'<l+1$ for some $l\in\Zb$,
  \item[(C3)] $z+z'+w+w'>0$, $|z+z'|<1$, $|w+w'|<1$.
  \end{enumerate}
  Then, as was shown by Borodin and Deift in \cite{borodin}, the determinant (\ref{f21det})
  is well-defined and $D(t)=\tau_{\scriptscriptstyle JMU}\left(t;\boldsymbol{\theta}\right)$ for
  the following choice of PVI parameters:
  \be\label{ourthetas}
  \boldsymbol{\theta}=(z+z'+w+w',z-z',0,w-w').
  \eb
  The original proof in \cite{borodin} that $D(t)$ satisfies Painlev\'e~VI is rather involved. In Section~\ref{twsection},
  we give an alternative simple derivation
  of this result in the spirit of \cite{twreview}.

  \begin{lem}
  Assume (C1)--(C3). Then the asymptotic expansion of $D(t)$ as $t\rightarrow0$
  has the form
  \be\label{IRas}
  D(t)=1-\kappa \cdot t^{1+z+z'+w+w'}+O\left(t^{2+z+z'+w+w}\right),
  \eb
  where
  \be\label{kappa}
  \kappa=\frac{\sin\pi z\sin\pi z'}{\pi^2}\,
  \Gamma\left[\begin{array}{c}1+z+w,1+z+w',1+z'+w,1+z'+w' \\ 2+z+z'+w+w',2+z+z'+w+w'\end{array}\right].
  \eb
  \end{lem}
  \pf
  As $t\rightarrow0$, one has $D(t)\sim 1-\int_{0}^{t}K(x,x)\,dx$. The result then follows from
  \ben
  A(x)\sim x^{\frac{z+z'+w+w'}{2}},\qquad B(x)\sim x^{\frac{z+z'+w+w'}{2}+1}\qquad\text{as }x\rightarrow0.
  \ebn
  Note that in the expression for $\kappa$ given in  Remark~7.2 in \cite{borodin} the gamma product
  is missing, which seems to be a typesetting error.
  \epf

  The asymptotics (\ref{IRas}) and $\sigma$PVI equation (\ref{zetapvi}) uniquely fix $D(t)$ by
  a result of~\cite{costin}. Gamma product in (\ref{kappa}) is a function of ${\theta}_{0}$, ${\theta}_{1}$,
  ${\theta}_{\infty}$ only, but $\displaystyle\text{\footnotesize{$\frac{\sin\pi z\sin\pi z'}{\pi^2}$}}$
  depends on an additional parameter (e.g. $z+z'$); hence we are dealing with a 1-parameter family
  of initial conditions.

  The results of \cite{borodin} can be extended to a larger set of parameters. This follows already from
  the observation that the subset of $\Cb^4$ defined by (C1)--(C3) is not stable under the transformations
  (S1)--(S3). However, instead of trying to identify all admissible values of $z,z',w,w'$, in the remainder of
  this paper we simply replace (C1)--(C3) by a much weaker (invariant) condition
  \begin{enumerate}
  \item[(C4)] $z+w,z+w',z'+w,z'+w'\notin\Zb_{<0}$ and $\mathrm{Re}\left(z+z'+w+w'\right)>0$,
  \end{enumerate}
  and \textit{define} $D(t)$ as the JMU $\tau$-function of Painlev\'e~VI with parameters (\ref{ourthetas}),
  whose leading behavior as $t\rightarrow0$ is specified by (\ref{IRas})--(\ref{kappa}). Our aim in the next
  sections is to determine the asymptotics of $D(t)$ as $t\rightarrow1$.

  \section{PBT $\tau$-function}\label{pbtsection} Palmer-Beatty-Tracy $\tau$-function \cite{lisovyy,beatty}
  is a regularized determinant of the quantum hamiltonian of a massive Dirac particle moving on
  the hyperbolic disk in the superposition of a uniform magnetic field $B$ and the field of two non-integer
  Aharonov-Bohm fluxes $2\pi\nu_{1,2}$ ($-1< \nu_{1,2}<0$)  located at the points $a_{1,2}$.

  Denote by $m$ and $E$ the particle mass and energy, by $-4/R^2$ the
  disk curvature and write $b=\frac{BR^2}{4}$, $\mu=\frac{\sqrt{(m^2-E^2)R^2+4b^2}}{2}$,
  $s=\tanh^2\frac{d(a_1,a_2)}{R}$, where $d(a_1,a_2)$ denotes the geodesic distance
  between $a_1$ and $a_2$. Then $\tau_{\scriptscriptstyle PBT}(s)$ can be expressed \cite{beatty}
  in terms of a solution $u(s)$ of the sixth Painlev\'e equation (\ref{pvi}):
  \begin{align}\label{taupbt1}
  \frac{d}{ds}\ln\tau_{\scriptscriptstyle PBT}(s)=&\,
  \frac{s(1-s)}{4u(1-u)(u-s)}\left(\frac{du}{ds}-\frac{1-u}{1-s}\right)^2\\
  \nonumber &\,-
  \frac{1-u}{1-s}\left(\frac{(\theta_{\infty}-1)^2}{4s}-\frac{(\theta_0+1)^2}{4u}+\frac{\theta_t^2}{4(u-s)}\right),
  \end{align}
  where the corresponding PVI parameters are given by
   \ben
  \boldsymbol{\theta}=(1+\nu_1+\nu_2-2b,0,2\mu,1+\nu_1-\nu_2).
  \ebn
  The initial conditions are specified by the asymptotics of $\tau_{\scriptscriptstyle PBT}(s)$
  as $s\rightarrow 1$, computed in \cite{lisovyy}:
  \be\label{taupbtas}
  \tau_{\scriptscriptstyle PBT}(s)=1-\kappa_{\scriptscriptstyle PBT}(1-s)^{1+2\mu}+O\left((1-s)^{2+2\mu}\right),
  \eb
  \ben
  \kappa_{\scriptscriptstyle PBT}=\frac{\sin\pi\nu_1\sin\pi\nu_2}{\pi^2}\,
  \Gamma\left[\begin{array}{c}2+\mu+\nu_1-b,\mu-\nu_1+b,2+\mu+\nu_2-b,\mu-\nu_2+b \\
  2+2\mu,2+2\mu\end{array}\right].
  \ebn

  Some resemblance between (\ref{zeta1}) and (\ref{taupbt1}) suggests that
  $\tau_{\scriptscriptstyle PBT}(s)$ is a special case of the JMU $\tau$-function.
  Indeed, consider the following transformation:
  \ben
  s\mapsto 1-t,\qquad u\mapsto \frac{1-t}{1-q}.
  \ebn
  In the notation of Table~1 of \cite{lisovyy_tykhyy}, this corresponds to B\"acklund transformation
  $r_xP_{xy}$ for Painle\-v\'e~VI. If $u(s)$ is a solution  with parameters
  $\boldsymbol{\theta}=(\theta_0,\theta_1,\theta_t,\theta_{\infty})$,
  then $q(t)$ solves PVI with parameters
  $\boldsymbol{\theta'}=(\theta_t,\theta_{\infty}-1,\theta_1,\theta_0+1)$.
  Straightforward calculation then shows that
  $\tau_{\scriptscriptstyle PBT}(1-t)=\tau_{\scriptscriptstyle JMU}(t;\boldsymbol{\theta'})$
  provided $\theta_1=0$.
  \begin{lem}\label{f21pbt}
  Under the following identification of parameters
  \begin{align}\label{pline1}
  z+z'+w+w'=2\mu,&\qquad z-z'=\nu_1-\nu_2,\qquad w-w'=2+\nu_1+\nu_2-2b,
  \\
  \label{pline2}
  &\!\!\!\!\!\cos\pi(z+z')=\cos\pi(\nu_1+\nu_2),
  \end{align}
  we have $D(t)=\tau_{\scriptscriptstyle PBT}(1-t)$.
  \end{lem}
  \pf
  It was shown above that if (\ref{pline1}) holds, then both $D(t)$ and
  $\tau_{\scriptscriptstyle PBT}(1-t)$ are JMU $\tau$-functions with the same $\boldsymbol{\theta}$.
  To show the equality, it suffices to verify that (\ref{pline2}) implies
  $\kappa=\kappa_{\scriptscriptstyle PBT}$.
  \epf

  Symmetries of $D(t)$ imply that $\tau_{\scriptscriptstyle PBT}(s)$ is invariant
  under transformations
  \begin{enumerate}
  \item[(S1)] $\mu\mapsto \mu$, $\nu_{1,2}\mapsto\nu_{1,2}$, $ b\mapsto 2+\nu_1+\nu_2-b$;
  \item[(S2)] $\mu\mapsto \mu$, $\nu_{1,2}\mapsto -2-\nu_{1,2}$, $b\mapsto -b$.
  \end{enumerate}
  These symmetries of $\tau_{\scriptscriptstyle PBT}(s)$ are by no means manifest,
  although they can also be deduced from the Fredholm determinant representation
  in \cite{lisovyy}, Theorem~1.1.

  \section{Painlev\'e VI from Tracy-Widom equations}\label{twsection}
    \subsection{Basic notation}
  Tracy and Widom \cite{twreview} have developed a systematic approach for deriving differential
  equations satisfied by Fredholm determinants of the form
  \be\label{det}
  D_I=\mathrm{det}\left(1-K_{I}\right),
  \eb
  where $K_I$ is an integral operator with the kernel
  \be\label{kernel}
  K_I(x,y)=\frac{\varphi(x)\psi(y)-\psi(x)\varphi(y)}{x-y},
  \eb
  on $L^2(J)$, with $J=\bigcup\limits_{j=1}^M\left(a_{2j-1},a_{2j}\right)$. The kernels of
  the form (\ref{kernel}) are called integrable and possess rather special properties:
  e.g. it was observed in \cite{IIKS} that the kernel of the resolvent $\left(1-K_I\right)^{-1}K_I$ is also integrable.

  The method of \cite{twreview} requires that $\varphi$, $\psi$ in (\ref{kernel}) obey a system of
  linear ODEs of the form
  \be\label{lpr}
  m(x) \frac{d}{dx}\left(\begin{array}{c}
  \varphi \\ \psi
  \end{array}\right)=\left(\sum\limits_{k=0}^{N}\mathcal{A}_kx^k\right)
  \left(\begin{array}{c}
  \varphi \\ \psi
  \end{array}\right),
  \eb
  where $m(x)$ is a polynomial and $\mathcal{A}_k\in\mathfrak{sl}_2(\Cb)$ ($k=0,\ldots,N$). 
  Note that a linear transformation
  $\left(\begin{array}{c} \varphi \\ \psi \end{array}\right)
  \mapsto G
  \left(\begin{array}{c} \varphi \\ \psi \end{array}\right)$ leaves $K_I(x,y)$ invariant
  provided $\mathrm{det}\,G=1$,
  and therefore $\left\{\mathcal{A}_k\right\}$ can be conjugated by an arbitrary $SL(2,\Cb)$-matrix.

  Our aim is to show that in the special case
  \ben
   m(x)=x(1-x),\qquad N=1, \qquad J=(0,t)
   \ebn
   the determinant (\ref{det})
   (i) coincides with the $_2F_1$ kernel determinant $D(t)$ and
   (ii) considered as a function of $t$, is a Painlev\'e~VI $\tau$-function.

  Let us temporarily switch to the notation of \cite{twreview} and introduce the  quantities
  \ben
  q(t)=\left[(1-K_I)^{-1}\varphi\right](t),\qquad  p(t)=\left[(1-K_I)^{-1}\psi\right](t),
  \ebn
  \ben
  u(t)=\langle \varphi |(1-K_I)^{-1}|\varphi\rangle,\qquad
   v(t)=\langle \varphi |(1-K_I)^{-1}|\psi\rangle,\qquad
    w(t)=\langle \psi |(1-K_I)^{-1}|\psi\rangle,
  \ebn
  where the inner products $\langle\,|\,\rangle$ are taken over $J$.
  Then
  \ben
  D_I^{-1}{D_I}'=qp'-pq',
  \ebn
  with primes denoting derivatives with respect to $t$. Tracy-Widom approach gives a system of
  nonlinear first order ODEs for $q$, $p$, $u$, $v$, $w$, which we are about to examine.
    \subsection{Derivation} Let $\mathcal{A}_1$ be diagonalizable, so that one can
  set
  \ben
  \mathcal{A}_0=\left(\begin{array}{cc}
  \alpha_0 & \beta_0 \\ -\gamma_0 & -\alpha_0
  \end{array}\right),\qquad
    \mathcal{A}_1=\left(\begin{array}{cc}
  \alpha_1 & 0 \\ 0 & -\alpha_1
  \end{array}\right).
  \ebn
  The Tracy-Widom equations then read
  \be\label{twe11}
  t(1-t)\left(\begin{array}{c}q' \\ p' \end{array}\right)
  =\left(\begin{array}{cc} \alpha & \beta \\
  -\gamma & -\alpha \end{array}\right)
  \left(\begin{array}{c}{q} \\ {p} \end{array}\right),
  \eb
  \be\label{twe12}
  u'=q^2,\qquad v'=pq,\qquad w'=p^2,
  \eb
  where
  \ben
  \alpha=\alpha_0+\alpha_1 t+v,\qquad \beta=\beta_0+(2\alpha_1-1)u,\qquad \gamma=\gamma_0-(2\alpha_1+1)w.
  \ebn
  The system (\ref{twe11})--(\ref{twe12}) has two first integrals
  \begin{align}
  \label{integral1} I_1=&\;2\alpha pq+\beta p^2+\gamma q^2-2\alpha_1 v,\\
  \label{integral2} I_2=&\;(v+\alpha_0)^2-\beta\gamma-2\alpha_1 t(1-t)pq+2\alpha_1(1-t)v-I_1 t.
  \end{align}

  Consider the logarithmic derivative
  $  \zeta(t)=t(t-1)D_I^{-1}{D_I}'$.
  It can be easily checked that
  \begin{align}
  \label{zeta}
  &\zeta=2\alpha pq+\beta p^2+\gamma q^2=2\alpha_1 v+I_1,\\
  \label{zetap}
  &\zeta'=2\alpha_1 pq,\\
  \label{zetapp}
  &t(1-t)\zeta''=2\alpha_1(\beta p^2-\gamma q^2).
  \end{align}
  Note that  $v$, $\alpha$ are expressible in terms of $\zeta$ and $pq$ in terms of $\zeta'$.
  Using (\ref{integral2}) and (\ref{zeta}) one may also write $\beta\gamma$ and $\beta p^2+\gamma q^2$
  in terms of $\zeta$ and $\zeta'$. Now squaring (\ref{zetapp}) we find a second order equation for $\zeta$:
  \be\label{zetaspvi}
  \bigl(t(1-t)\zeta''\bigr)^2+4\left(\zeta'-\alpha_1^2\right)\left(t\zeta'-\zeta\right)^2-
  4\zeta'\left(t\zeta'-\zeta\right)\left(\zeta'+2\alpha_0\alpha_1-I_1\right)=4(I_1+I_2)\left(\zeta'\right)^2.
  \eb

  If we parameterize the integrals $I_1$, $I_2$ as
  \begin{align*}
  I_1=&\;-k_1k_2+\alpha_1(2\alpha_0+\alpha_1),\\
  I_2=&\;\frac{(k_1+k_2)^2}{4}-\alpha_1(2\alpha_0+\alpha_1),
  \end{align*}
  and define
  \be\label{sigmavszeta}
  \sigma(t)=\zeta(t)-\alpha_1^2 t+\frac{\alpha_1^2+k_1 k_2}{2},\qquad\;
  \eb
  then (\ref{zetaspvi}) transforms into $\sigma$PVI equation (\ref{zetapvi})
  with parameters
  $\boldsymbol{\theta}=(k_1-k_2,k_1+k_2,0,2\alpha_1)$.
  Moreover, (\ref{sigmavszeta}) and the definition of $\zeta(t)$ imply that $D_I(t)$ coincides with
  the corresponding JMU $\tau$-function.

  The system (\ref{lpr}) has two linearly independent solutions, only one of which can be
  chosen to be regular as $x\rightarrow0$. This is the only solution of interest here, as if
  $\varphi$, $\psi$ have an irregular part, the operator $K_I$
  fails to be trace-class. The regularity further implies that  $q$, $p$, $u$, $v$, $w$ vanish as
  $t\rightarrow0$, and therefore the integrals $I_1$, $I_2$ are given by
  \ben
  I_1=0,\qquad I_2=\alpha_0^2-\beta_0\gamma_0.
  \ebn

  Choosing $\mathcal{A}_0$, $\mathcal{A}_1$ as above, one can still conjugate them by a diagonal matrix.
  Use this freedom to parameterize $\alpha_0$, $\beta_0$, $\gamma_0$, $\alpha_1$
  as follows:
  \begin{align*}
  \alpha_0=&\;-\frac{c}{2}-\frac{ab}{c-a-b},\\
   \beta_0=&\;-\frac{(c-a)(c-b)}{c-a-b},\\
  \gamma_0=&\;-\frac{ab}{c-a-b},\\
  \alpha_1=&\;\frac{c-a-b}{2},
  \end{align*}
  so that $I_2=\ds\frac{c^2}{4}$ and therefore $(k_1+k_2)^2=(a-b)^2$, $(k_1-k_2)^2=c^2$.
  Now if $\mathrm{Re}\,c>0$, the regular solution of (\ref{lpr}) is given by
  \be\label{lsreg}
  \left(\begin{array}{c}\varphi \\ \psi \end{array}\right)(x)=
  \mathrm{const}\cdot\left(\begin{array}{cc}
  1 & \frac{(c-a)(c-b)}{c(1+c)}\vspace{0.1cm} \\
  -1 & -\frac{ab}{c(1+c)}
  \end{array}\right)
  \left(\begin{array}{rl}
  x^{\frac{c}{2}}\left(1-x\right)^{-\frac{a+b}{2}}
  &{}_2F_1\left[\left.
  \begin{array}{c}a,b \\ c \end{array}\right|\ds \frac{x}{x-1}\right]\vspace{0.1cm} \\
  x^{1+\frac{c}{2}}\left(1-x\right)^{-1-\frac{a+b}{2}}
  &{}_2F_1\left[\left. \begin{array}{c}1+a,1+b \\ 2+c \end{array}
  \right|\ds \frac{x}{x-1}\right]
   \end{array}\right).
  \eb
   Setting $a=z+w'$, $b=z'+w'$, $c=z+z'+w+w'$ and comparing (\ref{lsreg})
   with (\ref{f21A})--(\ref{f21B}) we see that $K_I(t)$ coincides, up to an adjustable constant factor, with the $_2F_1$~kernel
   of Section~\ref{f21section}.

  \begin{rmk}
  A system similar to (\ref{twe11})--(\ref{twe12}) has already appeared in the Tracy-Widom analysis
  of the Jacobi kernel, see Section~V.C of \cite{twreview}.
  As the integral $I_2$ was not noticed there, the final result of \cite{twreview}
  was a \textit{third} order ODE (as one may well guess, it represents the first derivative of
  (\ref{zetaspvi}) in a disguised form). Later
  Haine and Semengue \cite{haine} have derived another third order equation for the Jacobi kernel determinant
  using the Virasoro approach of \cite{ASvM}, and obtained Painlev\'e~VI as the compatibility condition
  of the two equations. Our calculation gives, among other things, an elementary proof of this result.
  \end{rmk}

  \begin{rmk}
   For non-diagonalizable $\mathcal{A}_1$ it can be assumed that $ \mathcal{A}_1=\left(\begin{array}{cc}
  0 & 1 \\ 0 & 0
  \end{array}\right)$.
  The equations (\ref{twe12}) remain unchanged, whereas instead of (\ref{twe11}) we get
  \ben
  t(1-t)\left(\begin{array}{c}q' \\ p' \end{array}\right)
  =\left(\begin{array}{cc} \tilde{\alpha} & \tilde{\beta} \\
  -\tilde{\gamma} & -\tilde{\alpha} \end{array}\right)
  \left(\begin{array}{c}{q} \\ {p} \end{array}\right),
  \ebn
  where
  \ben
  \tilde{\alpha}=\alpha_0+v-w,\qquad \tilde{\beta}=\beta_0+s-u+2v,\qquad \tilde{\gamma}=\gamma_0-w.
  \ebn
   As before, we have two first integrals,
  \begin{align*}
  I_1=&\;2\tilde{\alpha} pq+\tilde{\beta} p^2+\tilde{\gamma} (q^2+1),\\
  I_2=&\;\tilde{\alpha}^2-\tilde{\beta}\tilde{\gamma}-t(1-t)p^2+(2t-1)\tilde{\gamma}-I_1t.
  \end{align*}
  The rest of the computation is completely analogous to the diagonalizable case.
  As a final result, one finds that the determinant $D(t)$ with $w=w'$
  is a $\tau$-function of Painlev\'e~VI with parameters $\boldsymbol{\theta}=(z+z'+2w,z-z',0,0)$.
  \end{rmk}

  \section{Jimbo's asymptotic formula}\label{jafsection} A remarkable result of
  Jimbo \cite{jimbo} relates the asymptotic behavior of the JMU $\tau$-function (\ref{jmutau})
  near the singular points $t=0,1,\infty$ to the monodromy of the associated linear system
  (\ref{fuchsian}).
  \begin{thmx}[Theorem~1.1 in \cite{jimbo}]\label{jaf}
  Assume that
  \begin{align}
  &\label{as1}\theta_0,\theta_1,\theta_t,\theta_{\infty}\notin\Zb,\tag{J1}\\
  &\label{as2}0\leq\mathrm{Re}\,\sigma_{0t}<1,\tag{J2}\\
  &\label{as3}\theta_0\pm\theta_t\pm\sigma_{0t},\theta_{\infty}\pm\theta_1\pm\sigma_{0t}\notin 2\Zb.\tag{J3}
  \end{align}
  Then $\tau_{\scriptscriptstyle JMU}(t)$ has the following asymptotic expansion
  as $t\rightarrow0$:
  \begin{align}
  \nonumber\tau_{\scriptscriptstyle JMU}(t)=\mathrm{const}\cdot t^{\frac{\sigma_{0t}^2-\theta_0^2-\theta_t^2}{4}}
  \Bigl[1&\,-\,\frac{\left(\theta_0^2-(\theta_t-\sigma_{0t})^2\right)
  \left(\theta_{\infty}^2-(\theta_1-\sigma_{0t})^2\right)}{16\sigma_{0t}^2(1+\sigma_{0t})^2}\,\hat{s}
  \;t^{1+\sigma_{0t}}\Bigr.\\
  \label{tjmuas1}&\,-\,\frac{\left(\theta_0^2-(\theta_t+\sigma_{0t})^2\right)
  \left(\theta_{\infty}^2-(\theta_1+\sigma_{0t})^2\right)}{16\sigma_{0t}^2(1-\sigma_{0t})^2}\,\hat{s}^{-1}
  \,t^{1-\sigma_{0t}}\\
  \nonumber
  &\,+\,\Bigl.\frac{(\theta_0^2-\theta_t^2-\sigma_{0t}^2)(\theta_{\infty}^2-\theta_1^2-
  \sigma_{0t}^2)}{8\sigma_{0t}^2}\,t\,+O\left(|t|^{2(1-\mathrm{Re}\,\sigma_{0t})}\right)\Bigr],
  \end{align}
  where $\sigma_{0t}\neq0$ and
  \ben
  \hat{s}=\Gamma\left[\begin{array}{c}
  1-\sigma_{0t},1-\sigma_{0t},1+\frac{\theta_0+\theta_t+\sigma_{0t}}{2},
  1-\frac{\theta_0-\theta_t-\sigma_{0t}}{2},1+\frac{\theta_{\infty}+\theta_1+\sigma_{0t}}{2},
  1-\frac{\theta_{\infty}-\theta_1-\sigma_{0t}}{2}\\
    1+\sigma_{0t},1+\sigma_{0t},1+\frac{\theta_0+\theta_t-\sigma_{0t}}{2},
    1-\frac{\theta_0-\theta_t+\sigma_{0t}}{2},1+\frac{\theta_{\infty}+\theta_1-\sigma_{0t}}{2},
    1-\frac{\theta_{\infty}-\theta_1+\sigma_{0t}}{2}
  \end{array}\right]s,
  \ebn
  \begin{align*}
  &\,s^{\pm1}\bigl(\cos\pi(\theta_t\mp\sigma_{0t})-\cos\pi\theta_0\bigr)
  \bigl(\cos\pi(\theta_1\mp\sigma_{0t})-\cos\pi\theta_{\infty}\bigr)=\\
  &\quad\,=\,\left(\pm\, i\sin\pi\sigma_{0t}\cos\pi\sigma_{1t}-\cos\pi\theta_t\cos\pi\theta_{\infty}
  -\cos\pi\theta_0\cos\pi\theta_1\right)e^{\pm i\pi\sigma_{0t}}\\
  &\quad\quad\;\;\;\pm i\sin\pi\sigma_{0t}\cos\pi\sigma_{01}+\cos\pi\theta_t\cos\pi\theta_1
  +\cos\pi\theta_{\infty}\cos\pi\theta_0.
  \end{align*}
  If $\sigma_{0t}=0$, then
  \begin{align*}
  \tau_{\scriptscriptstyle JMU}(t)=\mathrm{const}\cdot t^{-\frac{\theta_0^2+\theta_t^2}{4}}
  \Bigl[1&-\frac{\theta_1\theta_t}{2}\,t-
  \frac{(\theta_{\infty}^2-\theta_1^2)(\theta_0^2-\theta_t^2)}{16}\,t(\Omega^2+2\Omega+3)\Bigr.\\
  &+\frac{\theta_t(\theta_{\infty}^2-\theta_1^2)+\theta_1(\theta_0^2-\theta_t^2)}{4}\,
  t(\Omega+1)+o(|t|)\Bigr],
  \end{align*}
  where $\Omega=1-\hat{s}'-\ln t$ and
  \begin{align*}
  \hat{s}'=s'+\psi\left(1+\text{\footnotesize
  $\frac{\theta_0+\theta_t}{2}$}\right)+\psi\left(1+\text{\footnotesize
  $\frac{\theta_t-\theta_0}{2}$}\right)
  +\psi\left(1+\text{\footnotesize
  $\frac{\theta_{\infty}+\theta_1}{2}$}\right)+
   \psi\left(1+\text{\footnotesize
  $\frac{\theta_{1}-\theta_{\infty}}{2}$}\right)-4\psi(1).
  \end{align*}
  Here $\psi(x)$ denotes the digamma function and
  \ben
  s'=\,i\pi\,\frac{\cos\pi\sigma_{1t}+\cos\pi\sigma_{01}-
  \cos\pi\theta_0\, e^{i\pi\theta_1}-\cos\pi\theta_{\infty}\,e^{i\pi\theta_t}+i\sin\pi(\theta_1+\theta_t)
  }{\bigl(\cos\pi\theta_t-\cos\pi\theta_0\bigr)
  \bigl(\cos\pi\theta_1-\cos\pi\theta_{\infty}\bigr)}.
  \ebn
  \end{thmx}
  When one tries to determine from Theorem~\ref{jaf}
  the monodromy associated to the $_2F_1$ kernel solution $D(t)$
  of $\sigma$PVI, it turns out that all
  three assumptions (\ref{as1})--(\ref{as3}) are not satisfied:
  \begin{itemize}
  \item Firstly, (\ref{as1}) does not hold since in our case $\theta_t=0$. This requirement can nevertheless
  be relaxed as the appropriate non-resonancy condition for (\ref{fuchsian}) is $\theta_0,\theta_1,\theta_t,\theta_{\infty}\notin\Zb\backslash\{0\}$.
  The proof of
  asymptotic formulas when some $\theta$'s are
  equal to zero differs from that in \cite{jimbo} only in technical details;
  see e.g.~\cite{dubrovin}.
  \item If we blindly accept (\ref{tjmuas1}) then from $D(t\rightarrow0)\sim 1$ follows
  $\sigma_{0t}=\theta_0=z+z'+w+w'$. Thus (\ref{as2}) is violated unless $\mathrm{Re}\,\theta_0<1$
   and (\ref{as3}) does not hold in any case. Note, however, that (\ref{tjmuas1})
   admits a well-defined limit as $\theta_t=0$, $\sigma_{0t}\rightarrow\theta_0$. In this limit,
   the coefficients of $t$ and $t^{1-\sigma_{0t}}$ vanish; we also have
   \begin{align*}
   \cos\pi\sigma_{01}\rightarrow&\,\cos\pi\theta_{\infty}+\left(\cos\pi\theta_1-\cos\pi\sigma_{1t}\right)e^{-i\pi\theta_0},\\
   s(\theta_0-\sigma_{0t})\rightarrow&\,\frac{1}{\pi}\cdot
   \frac{\sin\pi\theta_0\,(\cos\pi\theta_1-\cos\pi\sigma_{1t})}{
   \sin\frac{\pi}{2}(\theta_{\infty}-\theta_0+\theta_1)\sin\frac{\pi}{2}(\theta_{\infty}+\theta_0-\theta_1)},
   \end{align*}
   and hence the coefficient of $t^{1+\sigma_{0t}}$ becomes
   \be\label{kappanaive}
   \frac{\cos\pi\theta_1-\cos\pi\sigma_{1t}}{2\pi^2}\,
   \Gamma\left[\begin{array}{c}
   1+\frac{\theta_0+\theta_1+\theta_{\infty}}{2},
   1+\frac{\theta_0+\theta_1-\theta_{\infty}}{2},
   1+\frac{\theta_0-\theta_1+\theta_{\infty}}{2},
   1+\frac{\theta_0-\theta_1-\theta_{\infty}}{2}\\
   2+\theta_0,2+\theta_0
   \end{array}\right].
   \eb
   \item Suppose that in our case the error estimate in (\ref{tjmuas1}) can be improved to $O\left(t^{2+\theta_0}\right)$
   (or at least to $o\left(t^{1+\theta_0}\right)$). Then, assuming that $0\leq\mathrm{Re}\left(z+z'\right)\leq1$ and comparing (\ref{kappanaive})
   with (\ref{kappa}), (\ref{ourthetas})  one would conclude that $\sigma_{1t}= z+z'$.
  \end{itemize}
  The above steps can indeed be justified --- after some tedious analysis
  going into the depths of Jimbo's proof. Alternatively, the monodromy
  can be extracted from Sections~3, 4 of \cite{borodin}, where $\sigma$PVI equation
  for $D(t)$ has itself been derived from a Riemann-Hilbert problem.

  \section{Asymptotics of $D(t)$ as $t\rightarrow1$}\label{mainsection}
  Once the monodromy is known, the asymptotics of $\tau_{\scriptscriptstyle JMU}(t)$ as $t\rightarrow1$
  can be determined from Jimbo's formula after substitutions
  $t\leftrightarrow1-t$, $\theta_0\leftrightarrow\theta_1$, $\sigma_{0t}\leftrightarrow\sigma_{1t}$,
  $\sigma_{01}\rightarrow\tilde{\sigma}_{01}$, where
  \be\label{jmissing}
  2\cos\pi\tilde{\sigma}_{01}=\mathrm{Tr}\left(M_0M_t^{-1}M_1M_t\right)=p_0p_1+p_tp_{\infty}-p_{0t}p_{1t}-p_{01}.
  \eb
  \begin{rmk} The transformation $\sigma_{01}\rightarrow\tilde{\sigma}_{01}$ is missing in \cite{jimbo}
  due to an incorrectly stated symmetry: the relation
  $\tau_{\scriptscriptstyle JMU}\left(1-t;
  M_0,M_t,M_{1}\right)=\mathrm{const}\cdot
  \tau_{\scriptscriptstyle JMU}\left(t; M_1,M_t,M_{0}\right) $
  on p.~1144 of \cite{jimbo} should be replaced by
  \ben
  \tau_{\scriptscriptstyle JMU}\left(1-t;
  M_0,M_t,M_{1}\right)=\mathrm{const}\cdot
  \tau_{\scriptscriptstyle JMU}\left(t;
  (M_t M_0)^{-1} M_1 M_t M_0,(M_0)^{-1}M_t M_0,M_{0}\right),
  \ebn
  which can be understood from Fig.~2.
 \begin{figure}[!h]
 \begin{center}
 \resizebox{12cm}{!}{
 \includegraphics{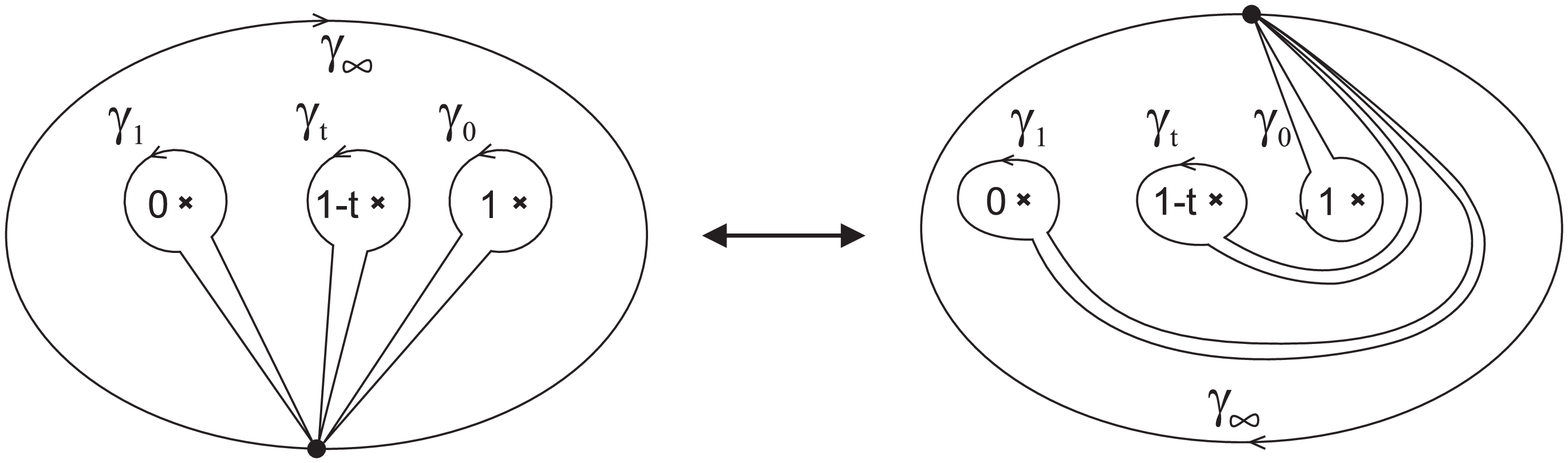}} \\
 Fig. 2: Homotopy basis after transformation
 $\lambda\mapsto1-\lambda$, $t\mapsto 1-t$.
 \end{center}
 \end{figure}
 \end{rmk}
  \begin{prop}\label{UVas}
  Assume that $0\leq\mathrm{Re}\left(z+z'\right)<1$ and
  \ben
  z,z',w,w',z+z'+w,z+z'+w'\notin\Zb.
  \ebn
  \begin{enumerate}
  \item If $z+z'\neq0$, then the following asymptotics is valid as $t\rightarrow1$:
  \begin{align}
  \nonumber D(t)=C\left(1-t\right)^{zz'}\Bigl[1&\,+\frac{zz'\left((z+z'+w)(z+z'+w')+ww'\right)}{(z+z')^2}\,(1-t)\Bigr.\\
  \label{dtas1} \Bigl.&\,-a^+(1-t)^{1+z+z'}-a^-(1-t)^{1-z-z'}+O\left((1-t)^{2-2\,\mathrm{Re}(z+z')}\right)
  \Bigr],
  \end{align}
  where $C$ is a constant and
  \ben
  a^{\pm}=\Gamma\left[\begin{array}{c}
  \mp z\mp z',\mp z\mp z',1\pm z,1\pm z',1+w+\frac{z+z'}{2}\pm \frac{z+z'}{2},1+w'+\frac{z+z'}{2}\pm \frac{z+z'}{2} \\
  2\pm z\pm z',2\pm z\pm z',\mp z,\mp z',w+\frac{z+z'}{2}\mp \frac{z+z'}{2},w'+\frac{z+z'}{2}\mp \frac{z+z'}{2}
  \end{array}\right].
  \ebn
  \item Similarly, if $z+z'=0$, then
  \begin{align}
  \nonumber D(t)=C(1-t)^{-z^2}\Bigl[1&\,+z^2ww'\,(1-t)(\tilde{\Omega}^2+2\tilde{\Omega}+3)\Bigr.\\
  \Bigl.&\,+z^2(w+w')
  \label{dtas2}(1-t)(\tilde{\Omega}+1)+o(1-t)\Bigr],
  \end{align}
  where $\tilde{\Omega}=1-a'-\ln (1-t)$ and
  \ben
  a'=\psi(1+z)+\psi(1-z)+\psi(1+w)+\psi(1+w')-4\psi(1).
  \ebn
  \end{enumerate}
  \end{prop}
  \pf
  Take into account that in our case $\theta_t=0$, $\sigma_{0t}=\theta_0$ and replace
  $\theta_0\leftrightarrow\theta_1$, $\sigma_{0t}\leftrightarrow\sigma_{1t}$,
  $\sigma_{01}\rightarrow\tilde{\sigma}_{01}$. Different
  quantities in Theorem~\ref{jaf} then transform into
  \ben
  s\rightarrow s_{1t}= 1\,,\qquad s'\rightarrow s_{1t}'=0,
  \ebn
  \ben
  \hat{s}\rightarrow\hat{s}_{1t}=\Gamma\left[\begin{array}{c}
  1-\sigma_{1t},  1-\sigma_{1t}, 1+\frac{\theta_1+\sigma_{1t}}{2}, 1-\frac{\theta_1-\sigma_{1t}}{2},
  1+\frac{\theta_0+\theta_{\infty}+\sigma_{1t}}{2}, 1+\frac{\theta_0-\theta_{\infty}+\sigma_{1t}}{2} \\
  1+\sigma_{1t},  1+\sigma_{1t}, 1+\frac{\theta_1-\sigma_{1t}}{2}, 1-\frac{\theta_1+\sigma_{1t}}{2},
  1+\frac{\theta_0+\theta_{\infty}-\sigma_{1t}}{2}, 1+\frac{\theta_0-\theta_{\infty}-\sigma_{1t}}{2}
  \end{array}\right],
  \ebn
 \ben
  \hat{s}'\rightarrow\hat{s}_{1t}'=
  \psi\left(1+\text{\footnotesize
  $\frac{\theta_1}{2}$}\right)+\psi\left(1-\text{\footnotesize
  $\frac{\theta_1}{2}$}\right)
  +\psi\left(1+\text{\footnotesize
  $\frac{\theta_0+\theta_{\infty}}{2}$}\right)+
   \psi\left(1+\text{\footnotesize
  $\frac{\theta_{0}-\theta_{\infty}}{2}$}\right)-4\psi(1).
 \ebn
  The statement now follows from $\sigma_{1t}=z+z'$ and (\ref{ourthetas}).
  \epf

  The constant $C$ in (\ref{dtas1})--(\ref{dtas2}) remains as yet undetermined.
  We will find an expression for it  using Lemma~\ref{f21pbt} and earlier results of Doyon \cite{doyon},
  who conjectured that for vanishing magnetic field
  $\tau_{\scriptscriptstyle PBT}(s)$ coincides with a correlation function of twist fields
  in the theory of free massive Dirac fermions on the hyperbolic disk. The asymptotics of
  $\tau_{\scriptscriptstyle PBT}(s)$ as $s\rightarrow0$ and $s\rightarrow1$ is then fixed, respectively,
  by conformal behavior of the correlator and its form factor expansion. The basic statement
  of \cite{doyon} (supported by numerics) is that there indeed exists a solution of the appropriate
  $\sigma$PVI equation which interpolates between the two asymptotics.

  Although the proof that the correlator of twist fields satisfies $\sigma$PVI has not yet been found,
  there are further confirmations of Doyon's hypothesis:  long-distance asymptotics (\ref{taupbtas}) with
  $b=0$  and the exponent $zz'$ in the short-distance power law (\ref{dtas1}) reproduce the
  conjectures of \cite{doyon}.

  The QFT analogy also implies that for real $z,z'\in(0,1)$ such that $0<z+z'<1$
  and $w'=w-z-z'$ (this corresponds to $b=0$) the constant $C$ in (\ref{dtas1})
  can be expressed in terms of vacuum expectation values
  of twist fields, which have been computed in \cite{doyon} (see also \cite{lz}).
  The resulting conjectural evaluation is:
   \be\label{dc0}
  C\bigl|_{w'=w-z-z'}=G\left[\begin{array}{c}
  1-z,1+z,1-z',1+z',1+w,1+w,1+z+z'+w,1-z-z'+w\\
  1-z-z',1+z+z',1+z+w,1-z+w,1+z'+w,1-z'+w
  \end{array}\right],
  \eb
  where
  $\ds G\left[\begin{array}{c}a_1,\ldots,a_m \\ b_1,\ldots,b_n\end{array}\right]=
  \frac{\prod_{k=1}^m G\left(a_k\right)}{\prod_{k=1}^n G\left(b_k\right)}$
  and $G(x)$ denotes the Barnes function:
  \ben
  G(x+1)=\left(2\pi \right)^{{x}/{2}}\exp\left\{\text{\footnotesize$\frac{\psi(1)x^2-x(x+1)}{2}$}\right\}
  \prod\limits_{n=1}^{\infty}\left[\left(1+\frac{x}{n}\right)^n \exp\left\{\text{\footnotesize$-x+\frac{x^2}{2n}$}\right\}\right].
  \ebn

  In spite of what one might expect, extension of the above approach to the
  case $b\neq0$ turns out to be rather complicated. However, the simple structure of
  (\ref{dc0}) and the symmetries of the $_2F_1$ kernel suggest the following:
  \begin{conj}\label{dysonconj}
  Under assumptions of Proposition~\ref{UVas},  the constant $C$ in the asymptotic
  expansions (\ref{dtas1}), (\ref{dtas2}) is given by
  \be\label{dc}
  C=G\left[\begin{array}{c}
  1-z,1+z,1-z',1+z',1+w,1+w',1+z+z'+w,1+z+z'+w'\\
  1-z-z',1+z+z',1+z+w,1+z+w',1+z'+w,1+z'+w'
  \end{array}\right].
  \eb
  \end{conj}
  \noindent The formula (\ref{dc}) is clearly compatible with (\ref{dc0}) and
  (S1)--(S2). It has been checked both numerically and analytically as described below.

  \section{Numerics}\label{nchecksection}
  \noindent To verify Conjecture~\ref{dysonconj}, one can proceed in the following way:
  \begin{enumerate}
  \item The solution of PVI associated to the $_2F_1$ kernel solution $D(t)$
  of $\sigma$PVI (uniquely determined by (\ref{IRas}), (\ref{kappa})) has the
  following asymptotic behavior as $t\rightarrow0$:
  \be\label{aswt1}
  q(t)=t-\lambda_0\, t^{1+z+z'+w+w'}+
  O(t^{2+z+z'+w+w'}),
  \eb
  \ben
  \lambda_0=\frac{(1+z+z'+w+w')^2}{(z+w)(z'+w)}\,\kappa.
  \ebn
  \item In fact one can show that in this case
  \begin{align}\label{aswt2}
  q(t)=t-&\,\lambda_0\, t^{1+z+z'+w+w'}(1-t)^{1+z-z'}
  {}_2F_1\left[\left.\begin{array}{c}
  z+w, 1+z+w' \\
  1+z+z'+w+w'
  \end{array}\right|t\right]^2 \\
  \nonumber +&\, O(t^{2+2(z+z'+w+w')}).
  \end{align}
  \item
  Use this asymptotics as initial condition and integrate the corresponding
  PVI equation numerically for some admissible choice of $\boldsymbol{\theta}$.
  It is then instructive to check Proposition~\ref{UVas} by verifying that
  for $0<\mathrm{Re}(z+z')<1$ the asymptotic
  expansion of $q(t)$ as $t\rightarrow1$ is given by
  \ben
  q(t)=1-\lambda_1\left(1-t\right)^{1-z-z'}+o\left((1-t)^{1-\mathrm{Re}(z+z')}\right),
  \ebn
  where
  \begin{align*}
  \lambda_1=&\,\Gamma\left[\begin{array}{c}
  z+ z',z+ z',1- z,1- z',w,1+w' \\
  1- z- z',1- z- z', z, z',z+z'+w,1+z+z'+w'
  \end{array}\right]=\\
  =&\,\frac{(1- z- z')^2 }{w
  \left(z+z'+w'\right)}\,a^{-}\,.
  \end{align*}
  Similarly, for $z+z'=0$ one has a logarithmic behavior,
  \ben
  q(t)=1+(1-t)\left[z^2\left(\tilde{\Omega}+w^{-1}-1\right)^2-1\right]+O\left((1-t)^2\ln^4(1-t)\right).
  \ebn
  \item Finally, use $q(t)$ and the initial condition
  $D(t)\sim 1$ as $t\rightarrow0$ to compute $D(t)$ from (\ref{zetatau}), (\ref{zeta1}).
  Looking at the asymptotics of $D(t)$
  as $t\rightarrow1$, one can numerically check the formula~(\ref{dc}) for $C$.
  \end{enumerate}
  \section{Special solutions check}\label{achecksection}
  For special choices of parameters and initial conditions
  Painle\-v\'e~VI equation can be solved explicitly. All explicit solutions found so far are
  either algebraic or  of Picard or Riccati type. Algebraic solutions have been classified in
  \cite{lisovyy_tykhyy}; up to parameter equivalence, their list consists of 3~continuous
  families and 45~exceptional solutions.

  It turns out that the parameters of exceptional algebraic solutions cannot be transformed to satisfy
  $_2F_1$ kernel constraints $p_0=p_{0t}$, $p_t=2$. Continuous families, however, do contain
  representatives verifying these conditions. Explicit computation of the corresponding $\tau$-functions
  provides a number of analytic tests of Conjecture~\ref{dysonconj}, some of which are presented below.
  Our notation for PVI B\"acklund transformations
  follows Table~1 in \cite{lisovyy_tykhyy}.

  \begin{eg}
  Painlev\'e VI equation with parameters $\boldsymbol{\theta}=(1,\theta_1,0,\theta_1)$
  is satisfied by
  \ben
  q(t)=1-\frac{(2\theta_1-1)-(2\theta_1+1)\sqrt{1-t}}{(2\theta_1-3)-(2\theta_1-1)\sqrt{1-t}}\,\sqrt{1-t}.
  \ebn
  This two-branch solution is obtained by applying
  B\"acklund transformation $s_{\delta}s_x s_y s_z s_{\delta} s_z s_{\delta} P_{xy}$
  to Solution~II in \cite{lisovyy_tykhyy} (set $\theta_a=1$, $\theta_b=\theta_1$).
   An explicit formula for the corresponding JMU $\tau$-function can be found
  from (\ref{zetatau}), (\ref{zeta1}):
  \ben
  \tau_{\scriptscriptstyle JMU}(t)=\left[\frac{2\left(1-t\right)^{1/4}}{1+\sqrt{1-t}}\right]^{\frac{1-4\theta_1^2}{4}}.
  \ebn
  Note that $\tau_{\scriptscriptstyle JMU}(t\rightarrow0)=1-\frac{1-4\theta_1^2}{128}\,t^2+O\left(t^3\right)$, and therefore
  $\tau_{\scriptscriptstyle JMU}(t)$ coincides with the hypergeometric kernel determinant $D(t)$ if we set
  $z=w=\frac{1+2\theta_1}{4}$, $z'=w'=\frac{1-2\theta_1}{4}$.

  The asymptotics of $\tau_{\scriptscriptstyle JMU}(t)$ as $t\rightarrow1$ has the form
  \ben
  \tau_{\scriptscriptstyle JMU}(t)= 2^{\frac{1-4\theta_1^2}{4}}\left(1-t\right)^{\frac{1-4\theta_1^2}{16}}\left(1+
  O\left(\sqrt{1-t}\right)\right),
  \ebn
  which implies that $C=2^{\frac{1-4\theta_1^2}{4}} $. To verify that this coincides
  with the expression
  \ben
  C=G\left[\begin{array}{c} \frac{3+2\theta_1}{4},\frac{3-2\theta_1}{4},\frac{5+2\theta_1}{4},\frac{5+2\theta_1}{4},
  \frac{5-2\theta_1}{4},\frac{5-2\theta_1}{4},\frac{7+2\theta_1}{4},\frac{7-2\theta_1}{4}, \vspace{0.1cm} \\ \frac{1}{2},\frac{3}{2},\frac{3}{2},\frac{3}{2},
  \frac{3+2\theta_1}{2},\frac{3-2\theta_1}{2}
  \end{array}\right].
  \ebn
  given by
  Conjecture~\ref{dysonconj}, one can use the recursion relation $G(z+1)=\Gamma(z) G(z)$,
  the duplication formulas (\ref{multig}), (\ref{multigamma}) for Barnes and gamma functions,
  and the  value of $G\left(\frac12\right)$  from Appendix~A.
  \end{eg}
  \begin{eg}
  Consider the rational curve
  \ben
  q=\frac{(s+1)(s-2)(5s^2+4)}{s(s-1)(5s^2-4)},\qquad t=\frac{(s+1)^2(s-2)}{(s-1)^2(s+2)}.
  \ebn
  It defines a three-branch solution of PVI with parameters $\boldsymbol{\theta}=(2,0,0,2/3)$,
  which can be obtained from Solution~III  in \cite{lisovyy_tykhyy} (with $\theta=0$) by
  the transformation $t_x=s_x s_{\delta} \left(s_ys_zs_{\infty}s_{\delta}\right)^2$.

  The associated
  $\tau$-function is given by
  \ben
  \tau_{\scriptscriptstyle JMU}(t(s))= \frac{3^{\frac{15}{8}}}{2^{\frac{25}{9}}}\cdot\frac{s\,(s+2)^{\frac89}}{(s+1)^{\frac{15}{8}}(s-1)^{\frac{7}{72}}},
  \ebn
  where the normalization constant is introduced for convenience.
  The map $t(s)$ bijectively maps the interval $(2,\infty)$ onto $(0,1)$. Choosing
  the corresponding solution branch one finds that
  \begin{align*}
  \tau_{\scriptscriptstyle JMU}(t\rightarrow0)=&\,\,1-\frac{16}{19683}\,t^3+
  O\left(t^4\right),\\
  \tau_{\scriptscriptstyle JMU}(t\rightarrow1)\sim &\,\,3^{\frac{15}{8}}\cdot
  2^{-\frac{17}{6}}\cdot\left(1-t\right)^{\frac{1}{36}}.
  \end{align*}

  First asymptotics implies that $\tau_{\scriptscriptstyle JMU}(t)$ coincides with $D(t)$ provided
  $z=z'=\frac{1}{6}$, $w=\frac{7}{6}$, $w'=\frac{1}{2}$. From the second asymptotics we obtain
  $C=3^{\frac{15}{8}}\cdot 2^{-\frac{17}{6}}$, whereas Conjecture~\ref{dysonconj} gives
  \ben
  C=G\left[\begin{array}{c} \frac{3}{2},\frac{5}{2},\frac{5}{6},\frac{5}{6},\frac{7}{6},\frac{7}{6},
  \frac{11}{6},\frac{13}{6} \vspace{0.1cm} \\ \frac{2}{3},\frac{4}{3},\frac{5}{3},\frac{5}{3},
  \frac{7}{3},\frac{7}{3}
  \end{array}\right].
  \ebn
  Equality of both expressions can be shown using the known evaluations of $G\left(\frac{k}{6}\right)$,
  $k=1\ldots5$, see~\cite{adamchik} or Appendix~A.
  \end{eg}
  \begin{eg} Applying the transformation $\left(s_{\delta}s_xs_y\right)^3s_zs_{\infty}s_{\delta}r_x$
  to Solution~IV in \cite{lisovyy_tykhyy} and setting
  $\theta=0$, one obtains a four-branch solution of PVI with  $\boldsymbol{\theta}=(1,1/2,0,1)$
  parameterized by
  \ben
  q=\frac{s(2-s)(5s^2-15s+12)}{(3-s)(3-2s)},\qquad t=\frac{s(2-s)^3}{3-2s}.
  \ebn
  The corresponding $\tau$-function has the form
  \ben
  \tau_{\scriptscriptstyle JMU}(t(s))=\frac{2^{\frac{5}{12}}}{3^{\frac{15}{16}}}\cdot
  \frac{(3-s)^{\frac{15}{16}}}{(2-s)^{\frac{5}{12}}(1-s)^{\frac{5}{48}}}.
  \ebn

  Choose the solution branch with $s\in(0,1)$. From the asymptotics $\tau_{\scriptscriptstyle JMU}(t\rightarrow0)=1+\frac{15}{2048}\,t^2+O\left(t^3\right)$ follows
  that $\tau_{\scriptscriptstyle JMU}(t)$ coincides with $D(t)$ provided $z=\frac{5}{12}$, $z'=-\frac{1}{12}$,
  $w=\frac56$, $w'=-\frac16$. Leading term in the asymptotic behavior of $\tau_{\scriptscriptstyle JMU}(t)$ as
  $t\rightarrow1$ is
  \ben
  \tau_{\scriptscriptstyle JMU}(t\rightarrow1)\sim 2^{\frac{25}{18}}\cdot 3^{-\frac{15}{16}}\cdot
  (1-t)^{-\frac{5}{144}},
  \ebn
  so that we have $C=2^{\frac{25}{18}}\cdot 3^{-\frac{15}{16}}$. On the other hand,
  Conjecture~\ref{dysonconj} implies that
  \ben
  C=G\left[\begin{array}{c} \frac{5}{6},\frac{7}{6},\frac{11}{6},\frac{13}{6},\frac{7}{12},\frac{11}{12},
  \frac{13}{12},\frac{17}{12} \vspace{0.1cm} \\ \frac{2}{3},\frac{4}{3},\frac{3}{4},\frac{5}{4},
  \frac{7}{4},\frac{9}{4}
  \end{array}\right].
  \ebn
  To prove that these expressions are equivalent, (i) use the multiplication formula (\ref{multig})
  with $n=2$ and
  $z=\frac{1}{12},\frac{5}{12}$ to compute
  $G\left(\frac{1}{12}\right)G\left(\frac{5}{12}\right)
  G\left(\frac{7}{12}\right)G\left(\frac{11}{12}\right)$
  and (ii) combine the resulting expression with the evaluations of
  $G\left(\frac{k}{4}\right)$, $G\left(\frac{k}{6}\right)$.
  \end{eg}
  \section{Limiting kernels}\label{wmsection}
  \subsection{Flat space limit: PVI $\rightarrow$ PV}
  The interpretation of $D(t)$ as a determinant of a Dirac operator (Section~\ref{pbtsection})
  suggests to consider the flat space limit $R\rightarrow\infty$.  This corresponds to the
  following scaling limit of the $_2F_1$ kernel:
  \ben
  w'\rightarrow+\infty, \qquad 1-t\sim\frac{s}{w'},\qquad s\in(0,\infty).
  \ebn
  In this limit, $D(t)$ transforms into the Fredholm determinant
  $D_L(s)=\mathrm{det}\left(1-K_L\bigl|_{(s,\infty)}\right)$ with the kernel
  \ben
  K_L(x,y)=\lim_{w'\rightarrow+\infty}\frac{1}{w'}\,K\left(1-\frac{x}{w'},1-\frac{y}{w'}\right)=
  \lambda_L\frac{A_L(x)B_L(y)-B_L(x)A_L(y)}{x-y},
  \ebn
  \ben
  A_L(x)=
  x^{-\frac12} W_{\frac12-\frac{z+z'+2w}{2},\frac{z-z'}{2}}(x),
  \qquad B_L(x)=x^{-\frac12} W_{-\frac12-\frac{z+z'+2w}{2},\frac{z-z'}{2}}(x),\ebn
  \ben
  \lambda_L=\frac{\sin\pi z\sin\pi z'}{\pi^2}\,\Gamma\left[1+z+w,1+z'+w\right],
  \ebn
  where $W_{\alpha,\beta}(x)$ denotes the Whittaker's function of the 2nd kind.
  $K_L(x,y)$ is the so-called Whittaker kernel (see e.g.~\cite{borodin2}), which
  plays the same role in the harmonic analysis on the infinite
  symmetric group as the $_2F_1$ kernel does for $U(\infty)$.

  The function  $\ds \sigma_L(s)=s\, \frac{d}{ds}\ln D_L(s)$ satisfies a Painlev\'e~V
  equation written in $\sigma$-form:
  \be\label{spainlevev}
  \bigl(s\,\sigma_{L}''\bigr)^2=\bigl(2\left(\sigma_L'\right)^2-(z+z'+2w+s)\sigma_L'+\sigma_L\bigr)^2
  -4\bigl(\sigma_L'\bigr)^2\bigl(\sigma_L'-z-w\bigr)\bigl(\sigma_L'-z'-w\bigr).
  \eb
  This can be shown by considering the appropriate limit of the $\sigma$PVI equation
  for $D(t)$. An initial condition for (\ref{spainlevev}) is provided by the asymptotics
  \ben
  D_L(s\rightarrow\infty)=1-\lambda_L\, e^{-s}s^{-z-z'-2w-2}\left(1+O\left(s^{-1}\right)\right).
  \ebn

  To link our notation with the one used in the PV part of Jimbo's paper \cite{jimbo}, we
  should set $\left(\theta_0,\theta_t,\theta_{\infty}\right)^{(V)}_{\text{Jimbo}}=\left(z'+w,-z-w,z-z'\right)$,
  which gives $D_L(s)=e^{\frac{(z+w)s}{2}}\tau^{(V)}_{\text{Jimbo}}(s)$. This in turn allows to obtain from
  Theorem~3.1 in \cite{jimbo} the asymptotics of $D_L(s)$ as $s\rightarrow0$:
  \begin{prop}\label{whker}
  Assume that $0\leq\mathrm{Re}\left(z+z'\right)<1$ and $z,z',w,z+z'+w\notin \Zb$.
  \begin{enumerate}
  \item If $z+z'\neq0$, then
  \begin{align*}
  D_L(s)=&\;C_L s^{zz'}\left[1+\frac{zz'(z+z'+2w)}{(z+z')^2}s-a^+_L s^{1+z+z'}-a^-_L s^{1-z-z'}+O\left(s^{2-2\,\mathrm{Re}(z+z')}\right)\right],
  \end{align*}
  with
  \ben
  a^{\pm}_L=\Gamma\left[\begin{array}{c}
  \mp z\mp z',\mp z\mp z',1\pm z,1\pm z',1+w+\frac{z+z'}{2}\pm \frac{z+z'}{2} \\
  2\pm z\pm z',2\pm z\pm z',\mp z,\mp z',w+\frac{z+z'}{2}\mp \frac{z+z'}{2}
  \end{array}\right].
  \ebn
  \item If $z+z'=0$, then
  \begin{align*}
  D_L(s)=&\;C_L s^{-z^2}\left[1+z^2ws\bigl(\tilde{\Omega}_L^2+2\tilde{\Omega}_L+3\bigr)+z^2s\bigl(\tilde{\Omega}_L+1\bigr)+o(s)\right],
  \end{align*}
  where $\tilde{\Omega}_L=1-a'_L-\ln s$ and $a'_L=\psi(1+z)+\psi(1-z)+\psi(1+w)-4\psi(1)$.
  \end{enumerate}
  \end{prop}
  Note that the same result is obtained by considering the formal limit
  of the leading terms in the asymptotics of $D(t)$. This further suggests
  an expression for constant $C_L$:
 \begin{conj}\label{dyson_conj2}
  Under assumptions of Proposition~\ref{whker}, we have
  \ben
  C_L=\lim\limits_{w'\rightarrow\infty}\left(w'\right)^{-zz'}C=G\left[\begin{array}{c}
  1-z,1+z,1-z',1+z',1+w,1+z+z'+w\\
  1-z-z',1+z+z',1+z+w,1+z'+w
  \end{array}\right].
  \ebn
  \end{conj}
  \subsection{Zero field limit: PV $\rightarrow$ PIII}
  Next we consider  the limit of vanishing magnetic field, $B\rightarrow0$.
  In terms of the parameters of the Whittaker kernel, this translates into
  \ben
  w\rightarrow +\infty,\qquad s\sim\frac{\xi}{w},\qquad \xi\in(0,\infty).
  \ebn
  The scaled kernel is given by
  \ben
  K_M(x,y)=\lim_{w\rightarrow+\infty}\frac{1}{w}\,K_L\left(\frac{x}{w},\frac{y}{w}\right)=
  \frac{\sin\pi z\sin\pi z'}{\pi^2}\cdot\frac{A_M(x)B_M(y)-B_M(x)A_M(y)}{x-y},
  \ebn
  \ben
  A_M(x)=2\sqrt{x}\,K_{z'-z+1}\left(2\sqrt{x}\right),\qquad B_M(x)=2\,K_{z'-z}\left(2\sqrt{x}\right),
  \ebn
  where $K_{\alpha}(x)$ is the Macdonald function.

  Denote $D_M(\xi)=\mathrm{det}\left(1-K_M\bigl|_{(s,\infty)}\right)$ and
  introduce $\ds\sigma_M(\xi)=\xi \frac{d}{d\xi}\ln D_M(\xi)$. Then $\sigma_M(\xi)$
  solves the $\sigma$-version of a particular Painlev\'e~III equation:
  \be\label{spainleveiii}
  \bigl(\xi\sigma_M''\bigr)^2=4\sigma_M'(\sigma_M'-1)(\sigma_M-\xi\sigma_M')+(z-z')^2\bigl(\sigma_M'\bigr)^2.
  \eb
  To match the notation in \cite{jimbo}, we have to set
  $\left(\theta_0,\theta_{\infty}\right)^{(III)}_{\text{Jimbo}}=\left(z'-z,z-z'\right)$,
  which gives $D_L(s)=e^{\xi}\tau^{(III)}_{\text{Jimbo}}(\xi)$.
  The appropriate initial condition for this $\sigma$PIII is given by
  \be\label{piiias}
  D_M(\xi\rightarrow\infty)=1-\frac{\sin\pi z\sin\pi z'}{4\pi}\cdot \frac{e^{-4\sqrt{\xi}}}{\sqrt{\xi}}
  \left(1+\frac{4(z-z')^2-3}{8\sqrt{\xi}}+O\left({\xi}^{-1}\right)\right).
  \eb

  The asymptotics of $D_M(\xi)$ as $\xi\rightarrow0$ can now be obtained from Theorem~3.2 in \cite{jimbo}:
  \begin{prop}\label{dmas}
  Assume that $0\leq\mathrm{Re}\left(z+z'\right)<1$ and $z,z'\notin\Zb$.
  \begin{enumerate}
  \item If $z+z'\neq0$, then
  \ben
  D_M(\xi\rightarrow0)=C_M\,\xi^{zz'}\left[1+\frac{2zz'}{(z+z')^2}\xi -
  a^+_M \xi^{1+z+z'}-a^-_M \xi^{1-z-z'}+O\left(\xi^{2-2\,\mathrm{Re}(z+z')}\right)\right],
  \ebn
  with
  $a^{\pm}_M=\Gamma\left[\begin{array}{c}
  \mp z\mp z',\mp z\mp z',1\pm z,1\pm z' \\
  2\pm z\pm z',2\pm z\pm z',\mp z,\mp z'
  \end{array}\right]$.
  \item If $z+z'=0$, then
  \ben
  D_M(\xi\rightarrow0)=C_M\,\xi^{-z^2}\left[1+z^2\xi\bigl(\tilde{\Omega}_M^2+2\tilde{\Omega}_M+3\bigr)+o(\xi)\right],
  \ebn
  where $\tilde{\Omega}_M=1-a'_M-\ln \xi$ and $a'_M=\psi(1+z)+\psi(1-z)-4\psi(1)$.
  \end{enumerate}
  \end{prop}
  Analogously to the above, we suggest a conjectural expression for $C_M$:
  \begin{conj}\label{dyson_conj3}
  Under assumptions of Proposition~\ref{dmas}, we have
  \ben C_M=\lim\limits_{w\rightarrow\infty}w^{-zz'}C_L=
  G\left[\begin{array}{c}
  1-z,1+z,1-z',1+z'\\
  1-z-z',1+z+z'
  \end{array}\right].
  \ebn
  \end{conj}
  \noindent {\bf Partial proof.}\quad
  This formula can in fact be proved for real $z=z'\in\left[0,\frac12\right)$, though in an indirect way.
   Consider the solution $\psi(r)$ of the radial sinh-Gordon equation
  \ben
  \frac{d^2\psi}{dr^2}+\frac{1}{r}\frac{d\psi}{dr}=\frac12 \sinh2\psi,
  \ebn
  satisfying the boundary condition $\psi(r,\nu)\sim 2\nu K_0(r)$ as $r\rightarrow+\infty$. Define the function
  \ben
  \tau(r,\nu)=\exp\left\{\frac12\int_{r}^{\infty} u\left[\sinh^2\psi(u,\nu)-
  \left(\frac{d\psi}{du}\right)^2\right]du\right\}.
  \ebn
  and consider the logarithmic derivative $\ds\tilde{\sigma}(\xi)=\xi\frac{d}{d\xi}\ln\tau(2\sqrt{\xi},\nu)$.
  It is straightforward to show that $\tilde{\sigma}(\xi)$ satisfies $\sigma$PIII equation (\ref{spainleveiii})
  with $z=z'$. Further, a little calculation shows that, as $r\rightarrow+\infty$,
  \ben
  \tau(r,\nu)=1-\pi\nu^2\,\frac{ e^{-2r}}{2 r}\left(1-\frac{3}{4r}+O\left(r^{-2}\right)\right).
  \ebn
  Comparing this asymptotics with (\ref{piiias}), we conclude
  that $D_M(\xi)\Bigl|_{z=z'}=\tau\Bigl(2\sqrt{\xi},\ds\pm\frac{\sin\pi z}{\pi}\Bigr)$.

  On the other hand, $\tau(r,\nu)=\tau_B^{-1}(r,\nu)$, where $\tau_B(r,\nu)$ is a special case
  of the bosonic 2-point tau function of Sato, Miwa and Jimbo, which can be represented as
  an infinite series of integrals (formulas (4.5.30)--(4.5.31) in \cite{smj} with
  $l_1=l_2$). By direct asymptotic analysis of this series, Tracy \cite{tracy_tau}
  has proved that for $\nu\in \left[0,\frac{1}{\pi}\right)$ it has the following behavior
  as $r\rightarrow0$:
  \ben
  \tau_B(r,\nu)=e^{\beta(\nu)}r^{-\alpha(\nu)}\left(1+o(1)\right),
  \ebn
  with
  \ben
  \alpha(\nu)= \frac{\sigma^2(\nu)}{2},\qquad\sigma(\nu)=\frac{2}{\pi}\arcsin\pi\nu,
  \ebn
  \ben
  \beta(\nu)= 3\alpha(\nu)\ln2+\frac12\ln(1-\pi^2\nu^2)-2\ln\cos\frac{\pi\sigma(\nu)}{2}-
  2\ln\left( G\left[\begin{array}{c}
  \frac12,\qquad \frac12\vspace{0.1cm}\\
  \frac{1+\sigma(\nu)}{2},\frac{1-\sigma(\nu)}{2}
  \end{array}\right]\right).
  \ebn

  From $\ds\nu=\pm\frac{\sin\pi z}{\pi}$ one readily obtains $\sigma^2=2\alpha=4z^2$. Thus,
  in order to show that $\beta(\nu)$ reproduces the conjectured expression for $C_M$
  with $z=z'$, it is sufficient to prove the identity
  \ben
  G\left[\begin{array}{c}
  1+z,1+z,1-z,1-z\\
  1+2z,1-2z
  \end{array}\right]=2^{-4 z^2}\cos\pi z\;
   G\left[\begin{array}{c}
  \frac12,\frac12,\frac12,\frac12\vspace{0.1cm}\\
  \frac{1}{2}+z,\frac{1}{2}+z,\frac{1}{2}-z,\frac{1}{2}-z
  \end{array}\right].
  \ebn
  This, however, is a simple consequence of the duplication formula for Barnes function
  and the known evaluation of $G\left(\frac12\right)$.
  \hfill$\square$\vspace{0.1cm}

  \section*{Appendix A}
  \noindent
  Multiplication formula for Barnes function \cite{vardi}:
  \begin{align}\label{multig}
  \tag{A.1}\ln G(nx)=&\,\,\left(\frac{n^2x^2}{2}-nx\right)\ln 2-\frac{(n-1)(nx-1)}{2}\ln 2\pi+
  \frac{5}{12}\ln n
  -\frac{n^2-1}{12}+\\
  \nonumber &\;+\left(n^2-1\right)\ln A+\sum_{j=0}^{n-1}\sum_{k=0}^{n-1}\;\ln\, G\left(x+\frac{j+k}{n}\right),
  \end{align}
  where $A=\exp\left(\frac{1}{12}-\zeta'(-1)\right)$ denotes Glaisher's constant.\vspace{0.2cm}\\
  Asymptotic expansion as $|z|\rightarrow\infty$, $\mathrm{arg}\,z\neq\pi$:
  \be\label{asg}
  \tag{A.2}\ln G(1+z)=\left(\frac{z^2}{2}-\frac{1}{12}\right)\ln z-\frac{3z^2}{4}
  +\frac{z}{2}\ln 2\pi-\ln A+\frac{1}{12}+O\left(\frac{1}{z^2}\right).
  \eb
  Special values (see, e.g. \cite{adamchik}):
  \begin{align*}
  \ln G\left(\frac12\right)=&\,\,
  \frac{\ln 2}{24}-\frac{\ln\pi}{4}-\frac{3}{2}\ln A +\frac18,\\
  \ln G\left(\frac13\right)=&\,\,\frac{\ln 3}{72}+\frac{\pi}{18\sqrt3}
  -\frac23\ln \Gamma\left(\frac13\right)-\frac{4}{3}\ln A-
  \frac{1}{12\pi\sqrt3}\,\psi'\left(\frac13\right)+\frac19,\\
  \ln G\left(\frac23\right)=&\,\,\frac{\ln 3}{72}+\frac{\pi}{18\sqrt3}
  -\frac13\ln \Gamma\left(\frac23\right)-\frac{4}{3}\ln A-
  \frac{1}{12\pi\sqrt3}\,\psi'\left(\frac23\right)+\frac19,\\
  \ln G\left(\frac16\right)=&\,-\frac{\ln 12}{144}+\frac{\pi}{20\sqrt3}
  -\frac56 \ln\Gamma\left(\frac16\right)-\frac56 \ln A-
  \frac{1}{40\pi\sqrt3}\,\psi'\left(\frac16\right)+\frac{5}{72},\\
  \ln G\left(\frac56\right)=&\,-\frac{\ln 12}{144}+\frac{\pi}{20\sqrt3}
  -\frac16 \ln\Gamma\left(\frac56\right)-\frac56 \ln A-
  \frac{1}{40\pi\sqrt3}\,\psi'\left(\frac56\right)+\frac{5}{72},\\
  \ln G\left(\frac14\right)=&\,-\frac{3}{4}\ln\Gamma\left(\frac14\right)-\frac98\ln A+
  \frac{3}{32}-\frac{K}{4\pi},\\
  \ln G\left(\frac34\right)=&\,-\frac{1}{4}\ln\Gamma\left(\frac34\right)-\frac98\ln A+
  \frac{3}{32}+\frac{K}{4\pi},
  \end{align*}
  where $K$ is Catalan's constant. \vspace{0.2cm}\\
  When checking Conjecture~\ref{dysonconj} with explicit examples, one
  also needs the relations
  \be\label{multigamma}
  \tag{A.3}\Gamma(nx)= (2\pi)^{-\frac{n-1}{2}}n^{nx-\frac12}\prod_{k=0}^{n-1}\Gamma\left(x+\frac{k}{n}\right),
  \qquad \psi'(x)+\psi'(1-x)=\frac{\pi^2}{\sin^2\pi x}.
  \eb


\begin{thebibliography}{1000}
 \bibitem{adamchik}
 V. S. Adamchik, \textit{On the Barnes function}, Proc. 2001 Int. Symp. Symbolic
 and Algebraic Computation, Academic Press, (2001), 15-–20.
 \bibitem{ASvM}
 M. Adler, T. Shiota, P. van Moerbeke, \textit{Random matrices, vertex operators and the
 Virasoro algebra}, Phys. Lett.~\textbf{A208}, (1995), 67--78.
 \bibitem{baik}
 J. Baik, R. Buckingham, J. DiFranco, \textit{Asymptotics of Tracy-Widom distributions
 and the total integral of a Painlev\'e II function},  Comm. Math. Phys.~\textbf{280},
 (2008), 463--497;  preprint \texttt{arXiv:0704.3636 [math.FA]}.
 \bibitem{borodin2}
 A. Borodin, \textit{Harmonic analysis on the infinite symmetric group and
 the Whittaker kernel}, St. Petersburg Math. J.~\textbf{12}, (2001), 733--759.
 \bibitem{BO5}
 A. Borodin, G. Olshanski, \textit{Harmonic analysis on the infinite-dimensional unitary
 group and determinantal point processes}, Ann. Math.~\textbf{161}, (2005), 1319--1422;
 preprint \texttt{math/0109194 [math.RT]}.
 \bibitem{borodin}
 A. Borodin, P. Deift, \textit{Fredholm determinants, Jimbo-Miwa-Ueno tau-functions,
 and representation theory}, Comm. Pure Appl. Math.~\textbf{55}, (2002), 1160--1230;
 preprint \texttt{math-ph/0111007}.
 \bibitem{costin}
 O. Costin, R. D. Costin, \textit{Asymptotic properties of a family of solutions
 of the Painlev\'e equation $P_{VI}$}, Int. Math. Res. Notices~\textbf{22}, (2002), 1167-1182;
 preprint \texttt{math/0202235 [math.CA]}.
 \bibitem{dik}
 P. Deift, A. Its, I. Krasovsky, \textit{Asymptotics of the Airy-kernel determinant},
 Comm. Math. Phys. ~\textbf{278}, (2008), 643--678; preprint \texttt{math/0609451 [math.FA]}.
 \bibitem{dikz}
 P. Deift, A. Its, I. Krasovsky, X. Zhou, \textit{The Widom-Dyson constant for the
 gap probability in random matrix theory},  J. Comput. Appl. Math.~\textbf{202}, (2007), 26--47;
 preprint \texttt{math/0601535 [math.FA]}.
 \bibitem{doyon}
 B. Doyon, \textit{Two-point correlation functions of scaling fields in the Dirac theory
 on the Poincar\'e disk}, Nucl. Phys. \textbf{B675}, (2003), 607--630; preprint
 \texttt{hep-th/0304190}.
 \bibitem{dubrovin}
 B. Dubrovin, M. Mazzocco, \textit{Monodromy of certain Painlev\'e~VI transcendents
 and reflection groups}, Inv. Math.~\textbf{141}, (2000), 55--147;
 preprint \texttt{math.AG/9806056}.
 \bibitem{dyson}
 F. J. Dyson, \textit{Fredholm determinants and inverse scattering problems},
 Comm. Math. Phys.~\textbf{47}, (1976), 171--183.
 \bibitem{ehrhardt}
 T. Ehrhardt, \textit{Dyson's constant in the asymptotics of the Fredholm determinant
 of the sine kernel},  Comm. Math. Phys.~\textbf{262}, (2006), 317--341;
 preprint \texttt{math/0401205 [math.FA]}.
 \bibitem{haine}
 L. Haine, J.-P. Semengue, \textit{The Jacobi polynomial ensemble and
 the Painlev\'e~VI equation}, J. Math. Phys.~\textbf{40}, (1999), 2117--2134.
 \bibitem{IIKS}
 A. R. Its, A. G. Izergin, V. E. Korepin, N. A. Slavnov,
 \textit{Differential equations for quantum correlation functions}, Int. J. Mod. Phys.~\textbf{B4},
 (1990), 1003--1037.
 \bibitem{jimbo}
 M. Jimbo, \textit{Monodromy problem and the boundary condition for some Painlev\'e equations},
 Publ. RIMS, Kyoto Univ.~\textbf{18}, (1982), 1137--1161.
 \bibitem{jmms}
 M. Jimbo, T. Miwa, Y. M\^ori, M. Sato, \textit{Density matrix of an impenetrable Bose gas
 and the fifth Painlev\'e transcendent}, Physica~\textbf{1D}, (1980), 80--158.
 \bibitem{jmu}
 M. Jimbo, T. Miwa, K. Ueno, \textit{Monodromy preserving deformations of linear
 ordinary differential equations with rational coefficients I}, Physica~\textbf{2D},
 (1981), 306--352.
 \bibitem{krasovsky}
 I. V. Krasovsky, \textit{Gap probability in the spectrum of random matrices and
 asymptotics of polynomials orthogonal on an arc of the unit circle},
 Int. Math. Res. Not.~\textbf{2004}, (2004), 1249--1272;
 preprint \texttt{math/0401258 [math.FA]}.
 \bibitem{lisovyy}
 O.~Lisovyy, \textit{On Painlev\'e VI transcendents related to the Dirac operator
 on the hyperbolic disk}, J.~Math. Phys.~\textbf{49}, (2008), 093507;
 preprint \texttt{arXiv:0710.5744~[math-ph]}.
 \bibitem{lisovyy_tykhyy}
 O. Lisovyy, Yu. Tykhyy, \textit{Algebraic solutions of the sixth Painlev\'e
 equation}, preprint \texttt{arXiv:0809.4873 [math.CA]}.
 \bibitem{lz}
 S. Lukyanov, A. B. Zamolodchikov, \textit{Exact expectation values of local fields in  quantum
 sine-Gordon model}, Nucl. Phys.~\textbf{B493}, (1997), 571--587; preprint \texttt{hep-th/9611238}.
 \bibitem{beatty}
 J. Palmer, M. Beatty, C. A. Tracy, \textit{Tau functions for the Dirac operator
 on the Poincar\'e disk}, Comm. Math. Phys.~\textbf{165}, (1994), 97--173; preprint
 \texttt{hep-th/9309017}.
 \bibitem{smj}
 M. Sato, T. Miwa, M. Jimbo, \textit{Holonomic quantum fields IV},
 Publ. RIMS, Kyoto Univ.  \textbf{15}, (1979), 871--972.
 \bibitem{tracy_tau}
 C. A. Tracy, \textit{Asymptotics of a $\tau$-function arising in the two-dimensional Ising
 model}, Comm. Math. Phys.~\textbf{142}, (1991), 297--311.
 \bibitem{twairy}
 C. A. Tracy, H. Widom, \textit{Level-spacing distributions and the Airy kernel},
 Comm. Math. Phys.~\textbf{159}, (1994), 151--174; preprint \texttt{hep-th/9211141}.
 \bibitem{twreview}
 C. A. Tracy, H. Widom, \textit{Fredholm determinants, differential equations and matrix models},
 Comm. Math. Phys.~\textbf{163}, (1994), 33--72 ; preprint \texttt{hep-th/9306042}.
 \bibitem{vardi}
 I. Vardi, \textit{Determinants of Laplacians and multiple gamma functions},
 SIAM J. Math. Anal.~\textbf{19}, (1988), 493-507.
 \bibitem{wmtb}
 T. T. Wu, B. M. McCoy, C. A. Tracy, E. Barouch, \textit{Spin-spin correlation functions for
 the two-dimensional Ising model: exact theory in the scaling region}, Phys. Rev.~\textbf{B13},
 (1976), 316--374.
 \end{thebibliography}
 \end{document}